\documentclass[sigconf]{acmart}
\AtBeginDocument{%
  }
\usepackage{balance}
\setcopyright{acmlicensed}
\copyrightyear{2025} 
\acmYear{2025} 
\setcopyright{acmlicensed}\acmConference[MM '25]{Proceedings of the 33rd
 ACM International Conference on Multimedia}{October 27--31, 2025}{Dublin,
 Ireland}
 \acmBooktitle{Proceedings of the 33rd ACM International Conference on
 Multimedia (MM '25), October 27--31, 2025, Dublin, Ireland}
 \acmDOI{10.1145/3746027.3755136}
 \acmISBN{979-8-4007-2035-2/2025/10}

\usepackage{xcolor}
\usepackage{bm}
\usepackage{multirow}
\usepackage[table]{xcolor} 
\usepackage{threeparttable}
\usepackage{etoolbox}

\begin{document}

\title{SemanticGarment: Semantic-Controlled Generation and Editing of 3D Gaussian Garments}

\newcommand{\sharedfootnote}{\footnotemark[1]}
\author{Ruiyan Wang}
\email{rywang0627@sjtu.edu.cn}
\affiliation{%
  \institution{Shanghai Jiao Tong University}
  \city{Shanghai}
  \country{China}
}

\author{Zhengxue Cheng}
\email{zxcheng@sjtu.edu.cn}
\affiliation{%
  \institution{Shanghai Jiao Tong University}
  \city{Shanghai}
  \country{China}
}
\authornote{Corresponding authors.}

\author{Zonghao Lin}
\email{ruianlzh@sjtu.edu.cn}
\affiliation{%
  \institution{Shanghai Jiao Tong University}
  \city{Shanghai}
  \country{China}
}

\author{Jun Ling}
\email{lingjun@sjtu.edu.cn}
\affiliation{%
  \institution{Shanghai Jiao Tong University}
  \city{Shanghai}
  \country{China}
}

\author{Yuzhou Liu}
\email{7zlyzz@sjtu.edu.cn}
\affiliation{%
  \institution{Shanghai Jiao Tong University}
  \city{Shanghai}
  \country{China}
}

\author{Yanru An}
\email{1847912805@sjtu.edu.cn}
\affiliation{%
  \institution{Shanghai Jiao Tong University}
  \city{Shanghai}
  \country{China}
}

\author{Rong Xie}
\email{xierong@sjtu.edu.cn}
\affiliation{%
  \institution{Shanghai Jiao Tong University}
  \city{Shanghai}
  \country{China}
}

\author{Li Song}
\email{song_li@sjtu.edu.cn}
\affiliation{
  \institution{Shanghai Jiao Tong University}
  \city{Shanghai}
  \country{China}
}
\authornotemark[1]

\renewcommand{\shortauthors}{Ruiyan Wang et al.}

\ccsdesc[300]{Computing methodologies~Point-based models}
\ccsdesc[500]{Information systems~Multimedia content creation}
\ccsdesc[100]{Computing methodologies~Animation}
\keywords{3D Gaussian Splatting, 3D multimodal generation, Animation, Clothing generation and editing}
\begin{abstract}
3D digital garment generation and editing play a pivotal role in fashion design, virtual try-on, and gaming. Traditional methods struggle to meet the growing demand due to technical complexity and high resource costs. 
Learning-based approaches offer faster, more diverse garment synthesis based on specific requirements and reduce human efforts and time costs.
However, they still face challenges such as inconsistent multi-view geometry or textures and heavy reliance on detailed garment topology and manual rigging.
We propose SemanticGarment, a 3D Gaussian-based method that realizes high-fidelity 3D garment generation from text or image prompts and supports semantic-based interactive editing for flexible user customization. To ensure multi-view consistency and garment fitting, we propose to leverage structural human priors for the generative model by introducing a 3D semantic clothing model, which initializes the geometry structure and lays the groundwork for view-consistent garment generation and editing. Without the need to regenerate or rely on existing mesh templates, our approach allows for rapid and diverse modifications to existing Gaussians, either globally or within a local region. To address the artifacts caused by self-occlusion for garment reconstruction based on single image, we develop a self-occlusion optimization strategy to mitigate holes and artifacts that arise when directly animating self-occluded garments. 
Project page: \url{https://semanticgarment.github.io/}.
\end{abstract}
\maketitle

\begin{figure}[!tb]
\setlength{\abovecaptionskip}{4pt}
\centering
    \includegraphics[width=8cm, height=9.3cm]{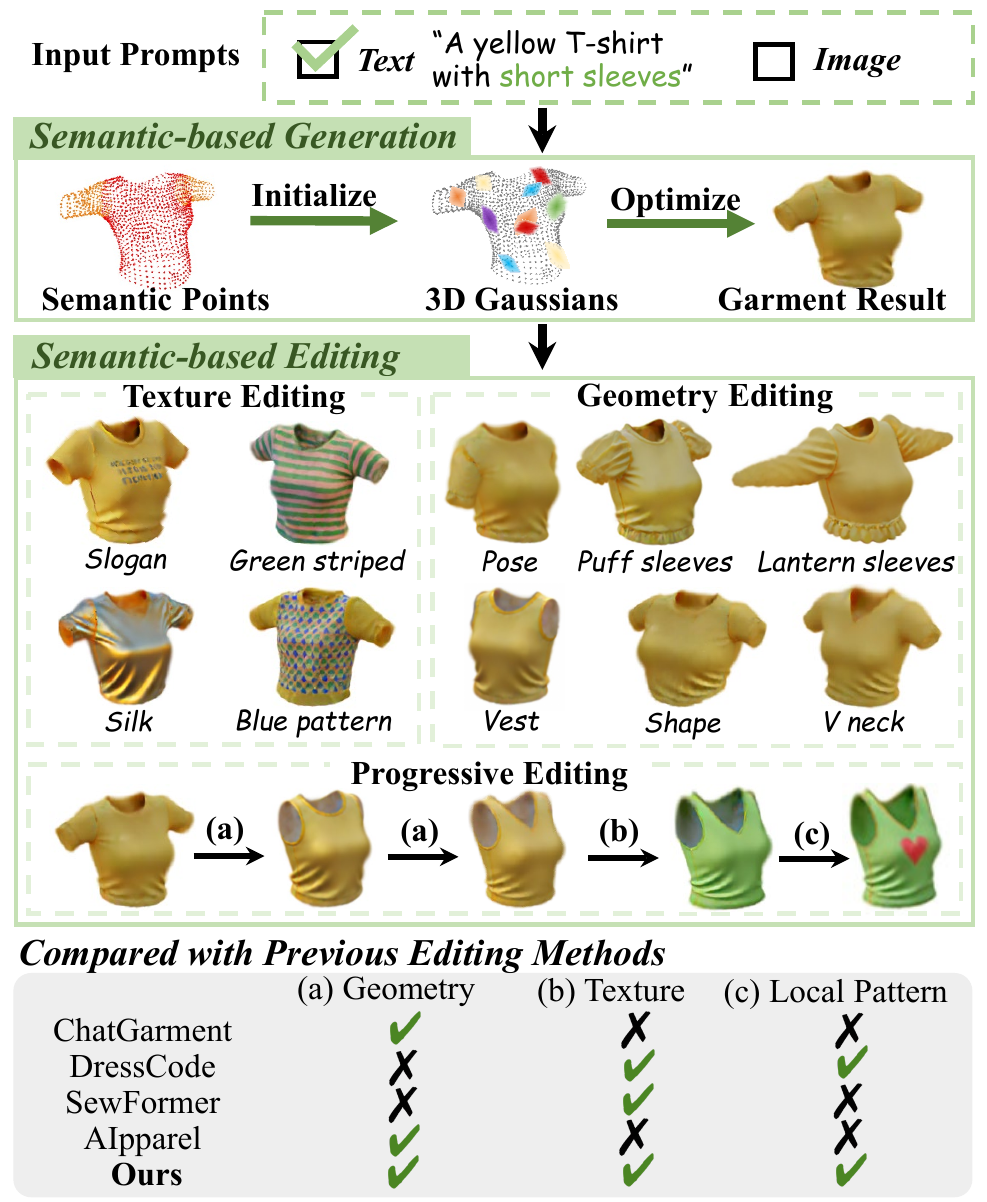}
    \caption{Our SemanticGarment generates high-quality, structurally consistent 3D garment Gaussians from text or real images, enabling diverse edits for fabric, style, pattern, pose, shape, and other user-defined requirements.}
\label{fig:Teaser}
\end{figure}


\section{Introduction}
\label{sec:intro}
The creation and control of 3D digital garment assets are crucial in various fields such as virtual reality (VR), fashion design, the film industry, and gaming. A key challenge in producing high-quality 3D garment assets lies in ensuring structural consistency and the ability to animate textures effectively.

Traditional 3D garment generation typically involves sketching, modeling, topology creation, texture mapping, and body binding, all of which require significant manual intervention.
In recent years, the rapid advancement of deep learning has driven notable progress in 3D garment generation using generative models, opening new possibilities for 3D clothing creation.
Learning-based clothing generation methods can be classified into two primary groups: direct garment generation from input prompts and generation guided by predefined sewing patterns.

The first category of methods~\cite{srivastava2024wordrobe, guo2025pgc,li2024garmentdreamer, liu2024clothedreamer,luo2024garverselod,srivastava2022xcloth,zhao2021learning,zhu2022registering,bhatnagar2019multi,qiu2023rec,sun2024if} focuses on directly generating the 3D geometry and texture of garments from image or text inputs. 
Due to the lack of strong multi-view supervision,  certain approaches~\cite{li2024garmentdreamer, liu2024clothedreamer} employ diffusion models to provide weak supervisory signals. While diffusion models can produce rich styles and detailed textures for garments, the absence of human body structural guidance often leads to issues such as multi-view geometric inconsistencies or garment deformations that fail to properly fit the human body.
The second category of methods~\cite{li2023isp,wang2018learning,shen2020gan,he2024dresscode, li2025dress, bian2024chatgarment, nakayama2024aipparel} employs predefined templates (sewing patterns) as intermediate representations to generate classical 3D garments in accordance with sewing patterns, thereby ensuring garment authenticity. However, these methods face challenges when modeling garments that deviate from the predefined templates or exhibit distinctive styles, such as puff sleeves. This is unless the sewing patterns are specifically modified for each unique garment.
Moreover, existing 3D garment generation methods~\cite{he2024dresscode, liu2023towards, bian2024chatgarment, nakayama2024aipparel} typically rely on predefined templates for further editing of the generated results. However, such editing scheme might bring about failure cases in local editing due to difficulties in accurately localizing the specified editing region.

We propose SemanticGarment, a 3D Gaussian-based approach for the fast generation and editing of diverse, well-fitting garments from text or image prompts. First, based on SMPL-X~\cite{pavlakos2019expressive}, we introduce a high-precision 3D semantic garment model, which provides human mesh with region-wise body structure semantics for different geometrical topologies of various garments. Building on this core model, we employ a semantic-based initialization method to support 3D assets through improving multi-view inconsistency and garment rigging. Since the SMPL-X mesh vertices are inherently sparse, they struggle to meet the rapid generation requirements. To improve generation efficiency, we design an interpolation densification strategy to effectively densify the initial positions of 3D Gaussians. During optimization, our method utilizes a pre-trained 2D diffusion model as guidance and applies SDS loss~\cite{poole2022dreamfusion} to obtain 3DGS garments with rich texture details and wrinkles. 
We also propose a semantic-based editing method based on the semantic model, allowing control over both global and local regions for geometric or texture modifications to meet diverse user needs. By identifying the target region, pruning and densifying the Gaussians, and applying SDS guidance, we effectively achieve the desired garment edits.

Our method supports garment generation from either textual descriptions or a single input image to produce 3D assets. However, existing image-based garment generation is affected by the self-occlusion problem. Self-occlusion refers to the phenomenon where certain regions of the garment are hidden by the body's pose or shape during 3D garment generation or animation, resulting in low-quality or artifacts. Reconstructed garments easily exhibit annoying appearance artifacts in occluded regions when they are animated by novel poses.
Prior solution, Garment3DGen~\cite{sarafianos2024garment3dgen} tries to address this issue through an automatic view selection algorithm, which iteratively generates textures from large to small regions by selecting views with the most unfilled pixels. However, this iterative process may terminate before full coverage is achieved, leading to incompleteness of occluded areas.
To this end, we propose a self-occlusion optimization method that refines both the position and color attributes of the synthesized Gaussians and optimizes the occluded regions to ensure structural integrity and accurate color representation. The resulting full garment can be seamlessly animated using Linear Blend Skinning (LBS)~\cite{loper2023smpl}, and facilitate the downstream applications such as virtual try-on.


Our contributions can be summarized as follows:

(1) We introduce a fine-grained 3D semantic clothing model based on human structural priors, enabling more detailed and effective garment generation and editing control.

(2) We propose a 3DGS-based clothing synthesis method that incorporates efficient human body semantic priors and implements bi-modal guidance by using images or text as input prompts.

(3) We present the first 3D Gaussian clothing editing method, supporting control over pattern, material, style, and shape.

(4) We design a self-occlusion optimization that refines Gaussians in self-occluded regions, effectively mitigating artifacts and voids while restoring realistic clothing shape and color.

\section{Related Work}
\label{sec:related_work}
\noindent\textbf{3D Content Generation.}
Recent research on 3D content generation can be categorized into two main types: inference-based methods and optimization-based methods. The inference-based methods~\cite{jun2023shap,nichol2022point,tang2024lgm,hong2023lrm, shen2024gamba, szymanowicz2024splatter} directly infer 3D structure and textures from prompts after training on large-scale datasets, enabling the efficient generation of 3D content in seconds. Representative method LGM~\cite{tang2024lgm} generates high-resolution 3D models from text or single-view images using multi-view Gaussian features as an efficient yet powerful representation. LRM~\cite{hong2023lrm}, Gamba~\cite{shen2024gamba}, and Splatter image~\cite{szymanowicz2024splatter} achieve photorealistic 3D content in single-image reconstruction tasks. 
The optimization-based methods~\cite{tang2023dreamgaussian,gaussiandreamer,li2024garmentdreamer, gao2023textdeformer} typically incorporate additional supervision, such as rich priors from 2D diffusion models, to optimize the details of 3D content.  DreamGaussian~\cite{tang2023dreamgaussian} significantly reduces the generation time of optimization-based 2D lifting methods and extracts the mesh to further enhance quality. GaussianDreamer~\cite{gaussiandreamer}, GarmentDreamer~\cite{li2024garmentdreamer}, and ClotheDreamer~\cite{liu2024clothedreamer} generate 3D assets by utilizing 3D pretrained diffusion models~\cite{jun2023shap,nichol2022point} or mesh priors, followed by optimization through the 2D diffusion model via score distillation.
\begin{figure*}[!tb]
\centering
    \includegraphics[width=1\linewidth]{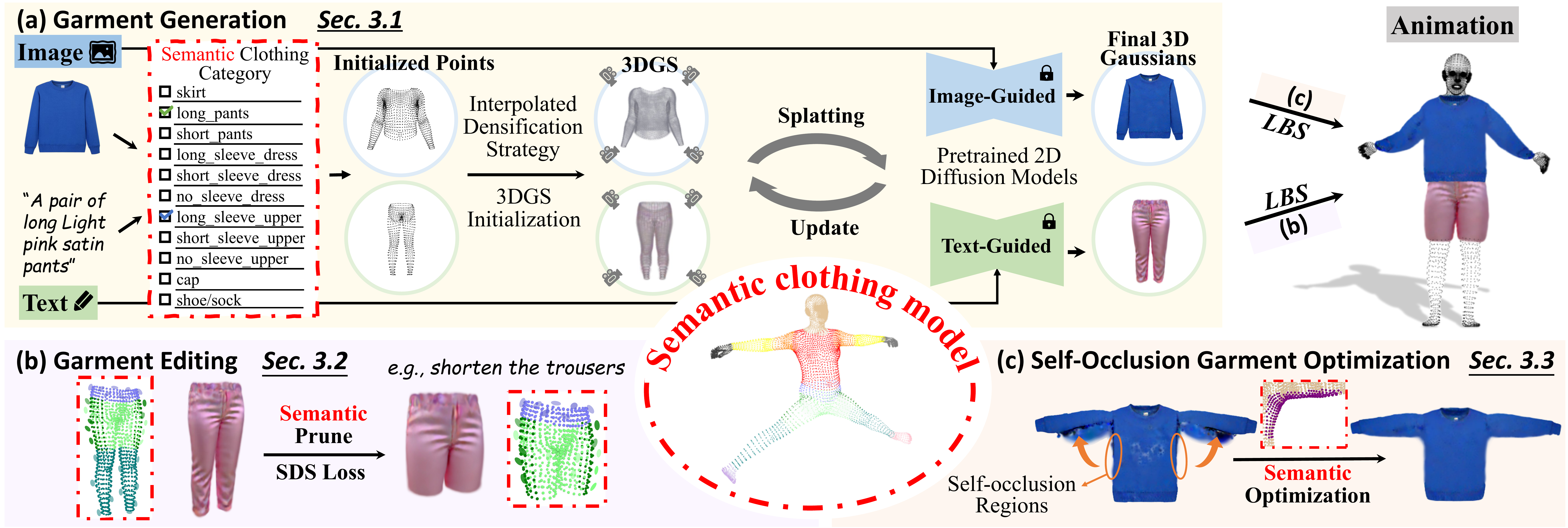}
    \caption{Method Overview. We introduce a core 3D semantic clothing model to enhance controllability in clothing generation, editing, and optimization. (a) Given a text or image prompt, our proposed generation approach leverages the semantic clothing model and an interpolated densification strategy to initialize 3DGS for different clothing types, while leveraging bi-modal pretrained 2D diffusion models to refine geometric and texture details. (b) Semantic-based editing can prune and densify the Gaussians according to user requirements, while applying SDS supervision to enhance details. (c) Self-occlusion optimization leverages semantic information to address the self-occlusion problem. }
\label{fig:method_pipeline}
\end{figure*}

\noindent\textbf{3D Garment Synthesis.}
Traditional 3D garments are created from 2D sewing patterns and simulated using physics-based simulations (PBS) to achieve realistic try-on effects, resulting in high labor and time costs~\cite{liu2024clothedreamer}. Consequently, learning-based methods have been proposed to automate 3D garment generation, enabling more flexible and diverse designs. Numerous studies~\cite{hong2022avatarclip, liao2024tada, jiang2023avatarcraft, cao2024dreamavatar, huang2024tech} focus on generating clothed avatars, allowing for the simultaneous generation of avatars and garments based on input prompts. 
AvatarCraft~\cite{jiang2023avatarcraft} and DreamAvatar~\cite{cao2024dreamavatar} generate high-quality 3D avatars by leveraging Text-to-Image diffusion models as guidance.
Another line of research~\cite{srivastava2024wordrobe, sarafianos2024garment3dgen, li2024garmentdreamer} focuses on generating stable and detailed garments independently from text or images, utilizing existing clothing templates as geometric priors. Given a base mesh and an image, Garment3DGen~\cite{sarafianos2024garment3dgen} proposes a novel deformation-based approach for generating garment geometry and texture. GarmentDreamer~\cite{li2024garmentdreamer} focuses on generating realistic and wearable garment meshes from text prompts, while WordRobe~\cite{srivastava2024wordrobe} enables text-driven 3D garment generation via a garment latent space. The reliance on clothing template topologies in these methods may limit garment shape diversity and hinder the generation of garments with specific shapes. In contrast, our method initializes independent garment shapes from the SMPL-X model, ensuring both adaptability to various avatar body shapes and flexibility in garment design.

\section{Method}
\label{sec:method}
We propose a high-fidelity clothing generation and editing framework based on 3D Gaussian Splatting~\cite{kerbl20233d}, supporting text or image guidance, as shown in Fig.~\ref{fig:method_pipeline}. 
To significantly improve the efficiency, we first initialize garment Gaussians from the 3D clothing semantic model in Sec.~\ref{sec:semantic_initialization}. Then we utilize pre-trained 2D diffusion models to add geometric and textural details in Sec.~\ref{sec:sds_guidance}. 
We thoroughly explain the editing process based on semantic information in Sec.~\ref{sec:edit}.
For self-occluding issue, we design the self-occlusion optimization in Sec.~\ref{sec:self-occlusion_optimize} to address holes and artifacts. 
\subsection{Semantic-based Generation}
\subsubsection{Semantic Initialization}
\label{sec:semantic_initialization}
\begin{figure}[!tb]
\setlength{\abovecaptionskip}{4pt} 
\centering
    \includegraphics[width=1\linewidth]{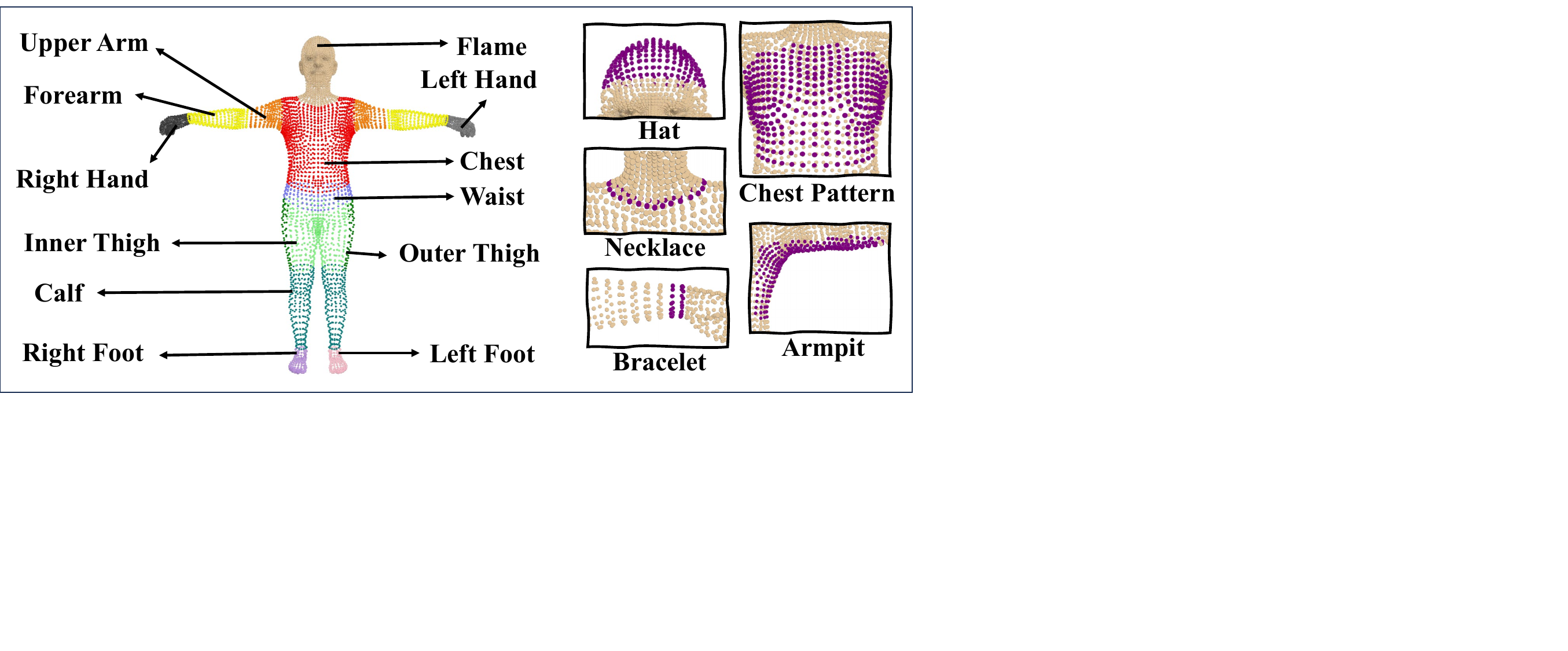}
    \caption{The visualization of the 3D semantic clothing model. By combining different semantic labels, any desired garment semantics can be obtained.}
\label{fig:semantic_model}
\end{figure}

\paragraph{3D Clothing Semantic Model.}
\label{sec:semantic_initialization_prior}
Existing studies~\cite{yang2024gaussianobject, zhu2024fsgs, fan2024instantsplat} show that in 3DGS representation, a suitable initial position enhances the generation of desired Gaussians, improving both efficiency and accuracy. To incorporate the body structure priors that clothing relies on, we introduce a novel semantic clothing model, as shown in Fig.~\ref{fig:semantic_model}, which encompasses semantic information of various clothing categories and accessories. Specifically, we annotate the SMPL-X model~\cite{pavlakos2019expressive} with different semantic labels according to body regions, allowing the initialization of diverse clothing types by combining these semantic attributes. SMPL-X is a parametric human body model for generating 3D human bodies. The generated mesh \bm$\mathcal{M}$ deforms based on shape parameters \bm$\beta$, pose parameters \bm$\theta$, and facial expression parameters \bm$\psi$. The linear blend skinning function~\cite{loper2023smpl}, \textit{LBS}($\cdot$), transforms the canonical template $T_P$ to the given pose $\bm\theta$ using skinning weights $\mathcal{W}$ and joint locations $J(\bm\beta)$ as:
\begin{equation}
\bm{\mathcal{M}}(\bm\beta, \bm\theta, \bm\psi) = LBS(T_P(\bm\beta, \bm\theta, \bm\psi), J(\bm\beta), \bm\theta, \mathcal{W}).
\label{Eq:3}
\end{equation}
Moreover, we also focus on key areas such as the ``chest pattern'' area and armpit area to facilitate subsequent clothing editing and optimization.
The initialization method based on SMPL-X semantic priors not only ensures that the garment conforms to the basic human structure from all perspectives (avoiding clothing flattening) but also facilitates subsequent editing and driven deformation to different postures.
\paragraph{Interpolated Densification Strategy.}
\label{sec:interpolated}
The inherent sparsity of the SMPL-X mesh vertices leads to longer optimization times when using Gaussian-based initialization. To improve efficiency, we propose an interpolated densification strategy leveraging the initial points. Specifically, we employ the KDTree~\cite{bentley1975multidimensional} algorithm to identify the nearest neighbor pairs for each initial point and perform uniform interpolation between each pair. 
\begin{figure}[!tb]
\setlength{\abovecaptionskip}{4pt} 
\centering
    \includegraphics[width=0.8\linewidth]{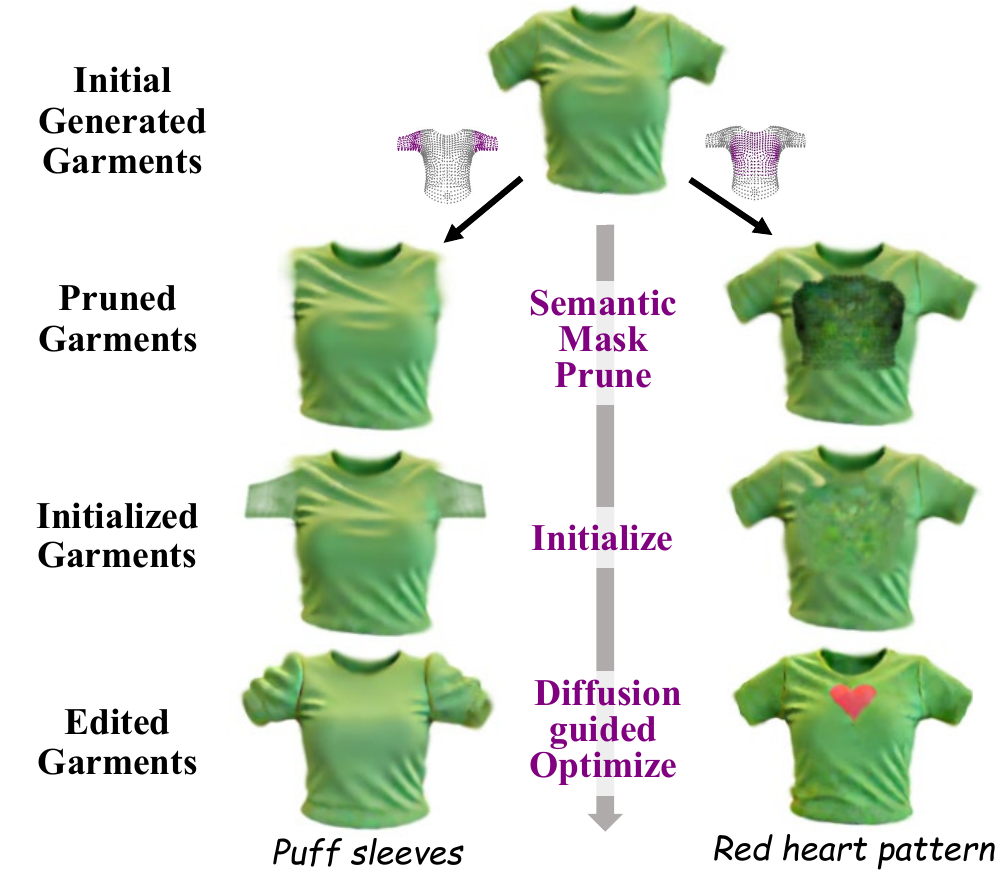}
    \caption{The visualization of local editing examples for pattern and style of the same garment.}
\label{fig:edit_example}
\end{figure}

\paragraph{3DGS Initialization.} 
In our method, each 3D Gaussian is characterized by central position \bm{$\mu$}, a covariance matrix \bm$\Sigma$, color \bm{$c$}, and opacity $\alpha$. To optimize the Gaussian parameters, the covariance matrix \bm{$\Sigma$} can be constructed from scale \bm{$s$} and a rotation quaternion \bm{$q$}.
The interpolated results are used as the position \bm{$\mu$} of initialized Gaussians. The initial color \bm{$c$} of the Gaussians is obtained through random sampling, with the scale \bm{$s$} determined by the distance between neighboring points, rotation \bm{$q$} initialized as a unit quaternion, and opacity $\alpha$ set to 0.1.
Existing research~\cite{hu2024gaussianavatar} indicates that, due to the imbalance of a single viewpoint, anisotropic 3D Gaussians tend to learn an inaccurate 3D shape that fits the most frequently observed view, leading to poorer performance in side views. Therefore, we set the scale values to be consistent across all dimensions to ensure the isotropy of the Gaussians.
\begin{figure}[!tb]
\setlength{\abovecaptionskip}{4pt} 
\centering
    \includegraphics[width=1\linewidth]{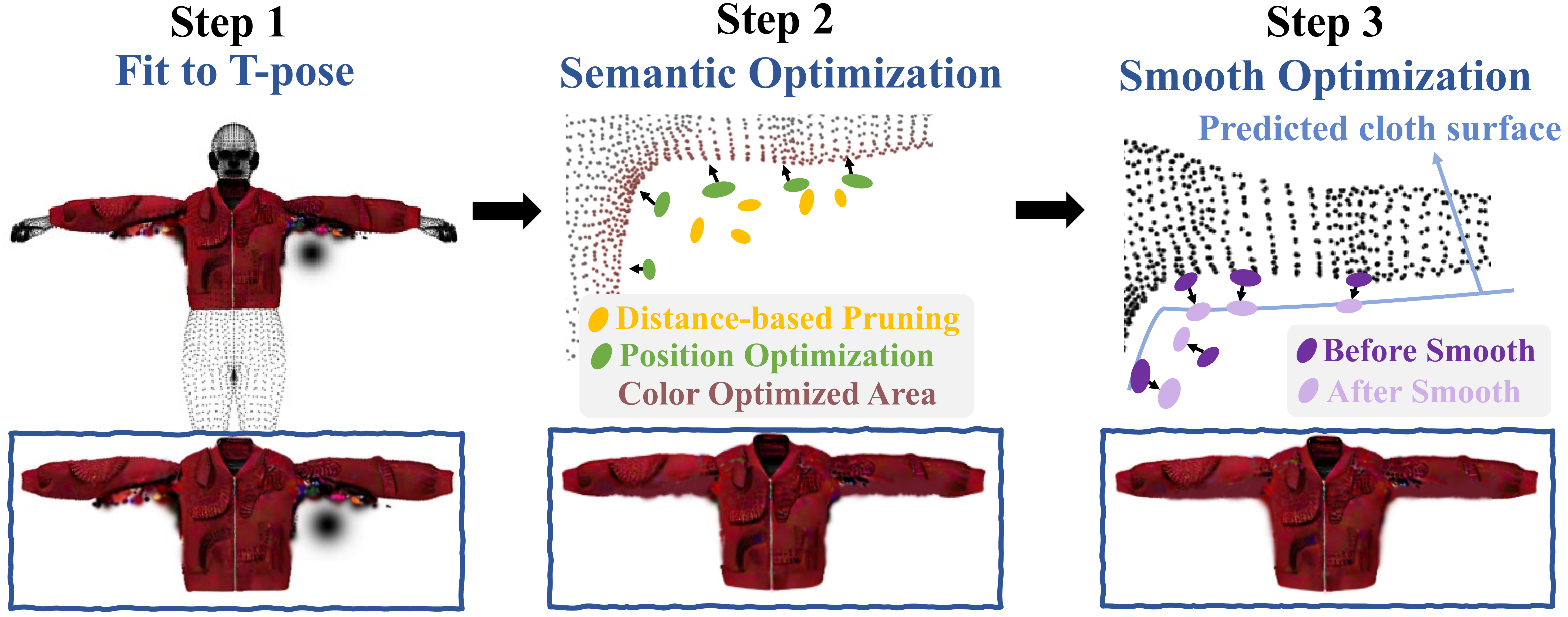}
    \caption{The Self-Occlusion Optimization process involves three key steps: (1) Fit to T-pose; (2) Semantic Optimization; (3) Smooth Optimization.}
\label{fig:method_Self-Occlusion}
\end{figure}
\subsubsection{Bi-modal 2D Diffusion Guidance}
\label{sec:sds_guidance}
To enhance the geometric and texture details of the garment, we utilize pre-trained 2D diffusion models for guidance after initialization, ensuring that the results closely align with the prompt.
Specifically, we render the initialized Gaussians into 2D images, and apply the Score Distillation Sampling (SDS) loss~\cite{poole2022dreamfusion} in Eq.~\ref{Eq:4} to optimize the 3DGS parameters $\Theta$.
SDS employs large-scale 2D pre-trained diffusion models as priors and refines the 3D content creation by minimizing the difference between the diffusion-sampled noise $\epsilon_\phi$ and the noise $\epsilon$ added to the rendered image $x$. 
\begin{equation}
\nabla_{\Theta} \mathcal{L}_{SDS} = \mathbb{E}_{t, \epsilon} \left[ w(t) \left(\hat\epsilon_\phi(x_{t}; y, t) - \epsilon \right) \frac{\partial x}{\partial \Theta} \right], 
\label{Eq:4}
\end{equation}
\noindent{where $t$ is the noise level, $x_t$ is the noisy version of $t$, $\hat{\epsilon}_\phi(x_{i}; y, t)$ is the predicted noise, and $w(t)$ is a weighting function.}
Our method supports text or image as input prompt $y$, with MVDream~\cite{shi2023mvdream} providing high-quality supervision for text and Stable-Zero123~\cite{liu2023zero} for image guidance.
If prompt $y$ is image, the RGB Loss $\mathcal{L}_{rgb}$ and the mask loss $\mathcal{L}_{mask}$ are also applied along with $\mathcal{L}_{SDS}$.
\begin{equation}
\mathcal{L}_{img} = \lambda_{rgb} \left\| I_{rgb} - \tilde{I}_{rgb} \right\|_2^2 + \lambda_{mask} \left\| I_{mask} - \tilde{I}_{mask} \right\|_2^2,
\end{equation}
\noindent{where $\lambda_{rgb}$ = $1\times10^5$, $\lambda_{mask}$ = 50, \( I \) and \( \tilde{I}\) are the input and generated images from the same view.}
\subsection{Semantic-based Editing}
\label{sec:edit}
\subsubsection{Global Editing}
\noindent\textbf{1) Texture editing.} We leverage the independence of Gaussian attributes to achieve diverse global editing. By applying 2D diffusion guidance to the color \textit{\textbf{c}} of the generated Gaussians and changing the text description related to material or color, we enable texture editing while preserving geometry.
\noindent\textbf{2) Shape editing.}
For diverse body shapes, we first compute the transformation matrix of the SMPL-X vertices between the standard shape and the target shape. The KNN algorithm binds the garment to the nearest mesh vertices, guiding the deformation of the Gaussians through vertex transformation, after which the garment can be deformed by applying the \textit{LBS} function in Eq.~\ref{Eq:3}.

\subsubsection{Local Editing}
Previous 3D editing works~\cite{park2023ed} applied loss within the masked pixels using 2D masks. However, differences in garment shape and viewpoint rotation can lead to inaccuracies or even invalidation of the masks.
Benefiting from the inherent 3D semantic information of generated garments, we can perform local editing tasks more effectively. 
Figure.~\ref{fig:edit_example} provides an intuitive illustration of the local pattern editing process. First, we identify the target region based on the editing requirements and utilize semantic information to prune the original Gaussians in the area. Subsequently, we rapidly initialize new Gaussians following the method described in Sec.~\ref{sec:semantic_initialization_prior}. To enhance editing efficiency, instead of randomly initializing the colors of the Gaussian points, we assign them the mean color value of the garment. Finally, we update the Gaussian attributes using the SDS loss to obtain the edited garment.

\subsection{Self-Occlusion Optimization}
\label{sec:self-occlusion_optimize}
The initialization in Sec.~\ref{sec:semantic_initialization} enables the generation of garments that fit the human body based on text prompts. However, when obtaining the corresponding garment Gaussian from real-world images, self-occlusion issues may arise.
As shown in Fig.~\ref{fig:method_Self-Occlusion}, the red jacket forms an ``A'' shape, with the inner side of the upper arm and the body sides falling within the self-occlusion regions. Directly applying the \textit{LBS} function to manipulate the garment leads to distorted Gaussians, as shown in ``Step 1''. To address this, we introduce the Self-Occlusion Optimization, a method that refines the unrealistic Gaussians and fills in the self-occluded regions through the introduction of specialized constraints.

\begin{table*}[!htbp]
\setlength{\abovecaptionskip}{4pt}
\renewcommand{\arraystretch}{0.85}
\caption{Quantitative comparison for clothing generation from text shows that our results achieve the highest text consistency and user preference. Our method achieves high fidelity in the shortest runtime among all optimization-based methods. }
\label{table:Text-3d}
\centering
\begin{tabular}{ccccccccc}
  \toprule
  \centering \multirow{2}*{Method} & \multirow{2}*{Type}& \multicolumn{3}{c}{CLIP-Score$\uparrow$}  & \multicolumn{3}{c}{User Study$\uparrow$} & \multirow{2}*{Run Time$\downarrow$}\\
  \cmidrule(r){3-5}  \cmidrule(r){6-8} 
   & & B/16 & B/32 & L/14  & 3D Consistency
 & Text Alignment
 & Image Quality
& \\
  \midrule
Shap-E~\cite{jun2023shap} & Inference & 30.178 & 29.818 & 25.767 & 3.600  & 2.709 & 3.338 & $\sim $5s \\
 LGM~\cite{tang2024lgm} & Inference & 32.332 & 31.803 & 27.887 & 3.338 &  3.300 & 3.114 &  \textbf{$ \sim $2.5s} \\
 3DGS~\cite{kerbl20233d}+SD2.1~\cite{rombach2022high} & Optimization & 30.062 & 29.279 & 25.633 & 2.575 & 3.063  & 2.100  & $ \sim $30mins \\
  
DreamGaussian~\cite{tang2023dreamgaussian} & Optimization & 27.985 & 26.815 & 24.126 & 3.338  & 3.101  &  2.329 & $ \sim $5mins \\
   
GaussianDreamer~\cite{gaussiandreamer} & Optimization & 29.929 & 28.958 & 26.374 & 4.000  & 2.949 & 3.638  & $ \sim $12mins \\

GarmentDreamer~\cite{li2024garmentdreamer}& Optimization & 28.940 & 27.666 & 24.306 & 3.165 & 2.775 & 3.388 & $ \sim $15mins\\

\textbf{Ours} & Optimization & \textbf{32.405} & \textbf{31.378} & \textbf{29.310} & \textbf{4.600} & \textbf{4.313} & \textbf{4.475}  & $ \sim $\textbf{4mins} \\
  \bottomrule
\end{tabular}
\end{table*}

\subsubsection{Fit to T-Pose}
For garments in a non-standard pose (i.e., not in the T-pose), each Gaussian point on the generated garment is bound to the human mesh, and the \textit{LBS} function is applied to transform the garment into the T-pose, enabling subsequent optimization.
\subsubsection{Semantic Optimization}
\label{sec:semantic_optimization}
\paragraph{Distance-based Pruning.} 
After \textit{LBS}, the originally occluded Gaussian points are moved away from the human body, and we prune them directly to improve optimization efficiency.
We first leverage the semantic prior from the SMPL-X to identify the self-occlusion region. 
The K-Nearest Neighbors (KNN) algorithm~\cite{cover1967nearest} is then performed to determine the nearest distances between Gaussian points and the SMPL-X vertices in the self-occluded region. 
By sorting the distances, we prune the redundant Gaussian points that are clearly distant from the human mesh.

\begin{table*}[!htbp]
\renewcommand{\arraystretch}{0.85}
\setlength{\abovecaptionskip}{4pt}
\caption{Quantitative Comparison for clothing guided by dataset images on DeepFashion3D v2~\cite{zhu2020deep} and ClothesNet~\cite{zhou2023clothesnet}.}
\label{table: Image-3d}
\centering
\begin{tabular}{cccccccc}
  \toprule
  \centering \multirow{2}*{Method} & \multicolumn{3}{c}{Image Quality}  & \multicolumn{2}{c}{User Study$\uparrow$} & \multicolumn{1}{c}{Image Similarity}\\
  \cmidrule(r){2-4}  \cmidrule(r){5-6}  \cmidrule(r){7-7}
    & PSNR$\uparrow$ & SSIM$\uparrow$ & LPIPS$\downarrow$  & 3D Consistency
 & Image Quality & CLIP-L/14$\uparrow$ 
\\
  \midrule
 Gamba~\cite{shen2024gamba} & 18.608 & 0.869 & 0.177 & 2.329  & 2.450 & 0.803\\

 LGM~\cite{tang2024lgm} & 19.012 & 0.874 & 0.153 & 3.125  & 3.300  & 0.827 \\
  
  DreamGaussian~\cite{tang2023dreamgaussian} & 19.563 & 0.889 & 0.132 & 2.125 & 1.863 & 0.833 \\
   
  TriplaneGaussian~\cite{zou2024triplane} & 19.890 & \textbf{0.898} & 0.148 &  2.738 & 3.438 & 0.828   \\

  \textbf{Ours}  & \textbf{20.686} & 0.891 & \textbf{0.127} & \textbf{4.563} & \textbf{4.413} & \textbf{0.835}   \\
  
  \bottomrule
\end{tabular}
\end{table*}
\paragraph{Position Optimization.}
After distance-based pruning, the remaining points do not fully conform to the garment's geometry. Therefore, we introduce a position loss function, denoted as $\mathcal{L}_{position}$. This method not only constrains the Gaussian points that are far from the human body surface, bringing them closer, but also effectively fills in gaps. Specifically, we calculate the sum of the closest distances $D_{total}$, between the Gaussians and SMPL-X vertices in the self-occlusion regions before optimization using the KNN algorithm. The optimization process halts when $D = 0.5D_{total}$, aiming to prevent the Gaussian points from adhering too closely to the body surface, thus preserving the garment's original geometric shape. This position optimization can be formulated as follows:
\begin{equation}
\mathcal{L}_{position} = \sum_{i=1}^{P} \| D_i \|_2^2  = \sum_{i=1}^{P} \| p_i - v_i \|_2^2,
\end{equation}
\noindent{where \(D_i\) is the distance from the \(i\)-th Gaussian position $p_i$ to its nearest vertex $v_i$, $P$ is the total number of Gaussians in the self-occlusion region.}
\paragraph{Color Optimization.} 
To address the issue of color inhomogeneity in self-occluded regions, we calculate the average color of the Gaussian points in the self-occlusion region as the target color $\bm{C}_{target}$, and then compute the $\mathcal{L}_2$ loss between the color of the Gaussian points and $\bm{C}_{target}$. This process can be formulated as:
\begin{equation}
\mathcal{L}_{color} = \sum_{i=1}^{P} \| \bm{C}_i - \bm{C}_{\text{target}} \|_2^2,
\end{equation}
\noindent{where $\bm{C}_i$ is the color of the $i$-th Gaussian.}
\subsubsection{Smooth Optimization}
Semantic optimization effectively constrains the position and color of the Gaussians but does not ensure the smoothness of the cloth surface (resulting in uneven bulges of the Gaussians).
In this optimization, we calculate the average distance $D_{avg}$ of the Gaussians to their nearest vertices after semantic optimization, and apply the smooth constraint $\mathcal{L}_{smooth}$ in Eq.~\ref{eq:8} to Gaussian points with distances $D > D_{avg}$. 
\begin{equation}
\mathcal{L}_{smooth} = \sum_{i=1}^{N} \mathbb{I}(D_i > D_{avg}) \| D_i - D_{avg} \|_2^2,
\label{eq:8}
\end{equation}
\noindent{where \(\mathbb{I}(D_i > D_{avg})\) is an indicator function that equals 1 if \(D_i > D_{avg}\), and 0 otherwise. }

\section{Experiments}
\label{sec:experiments}
\subsection{Implementation Details}
\begin{figure*}[!tb]
\setlength{\abovecaptionskip}{4pt} 
\centering
    \includegraphics[width=1\linewidth]{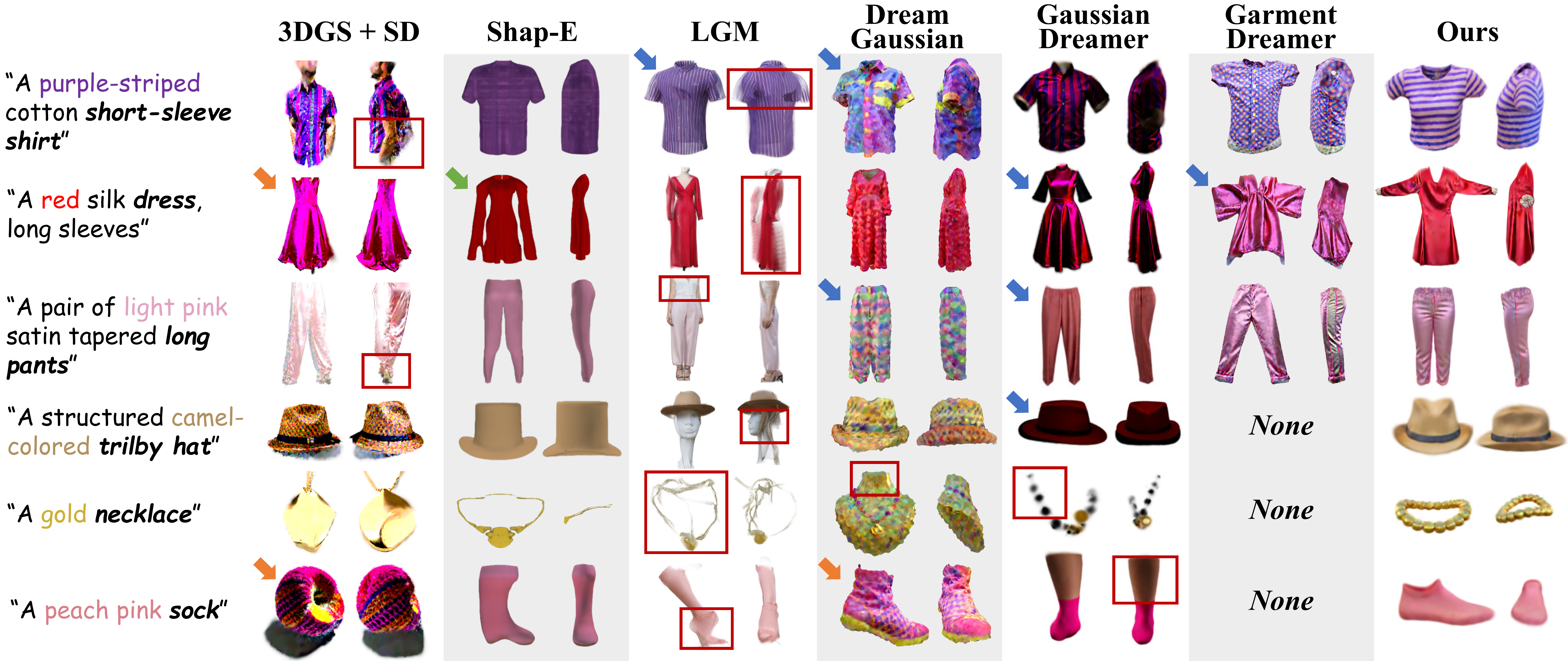}
    \caption{Comparisons on text-guided generative garments. The ``None'' label in GarmentDreamer results indicates no accessory templates. Results in the red box show ambiguity or extra content (e.g., human body). Green, blue, and orange arrows highlight incorrect material, color/pattern, and garment type, respectively. Our method generates garments with 3D consistency and high alignment with the text prompts. }
\label{fig:Text-3d}
\end{figure*}
\begin{figure}[!tb]
\setlength{\abovecaptionskip}{4pt} 
\centering
    \includegraphics[width=1\linewidth]{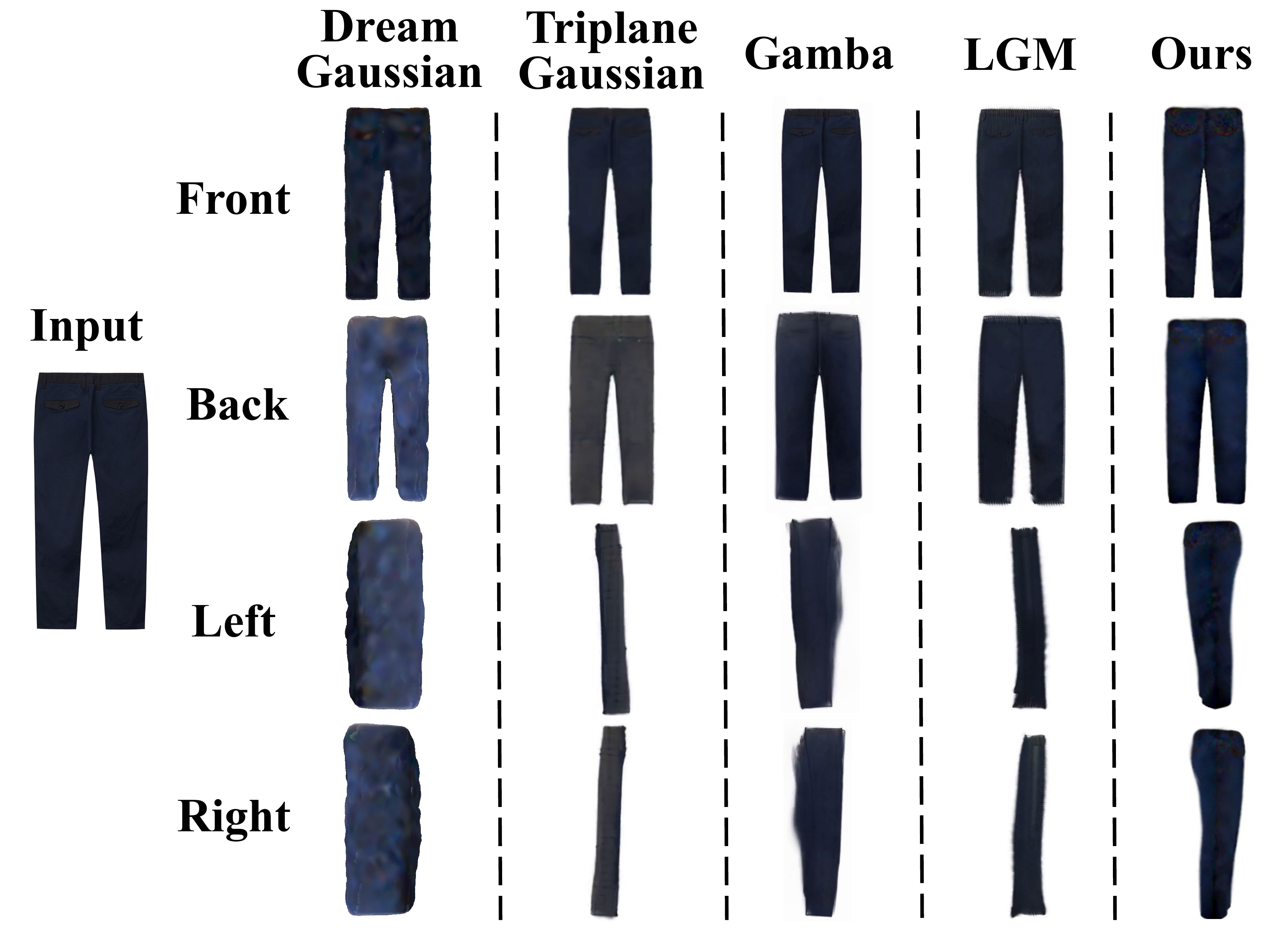}
    \caption{Qualitative comparisons on image-guided generative garments. The results demonstrate that our method generates clothing that aligns more accurately with the human body structure from multiple views.}
\label{fig:Image-3d}
\end{figure}
\begin{figure}[!tb]
\setlength{\abovecaptionskip}{4pt} 
\centering
    \includegraphics[width=1\linewidth]{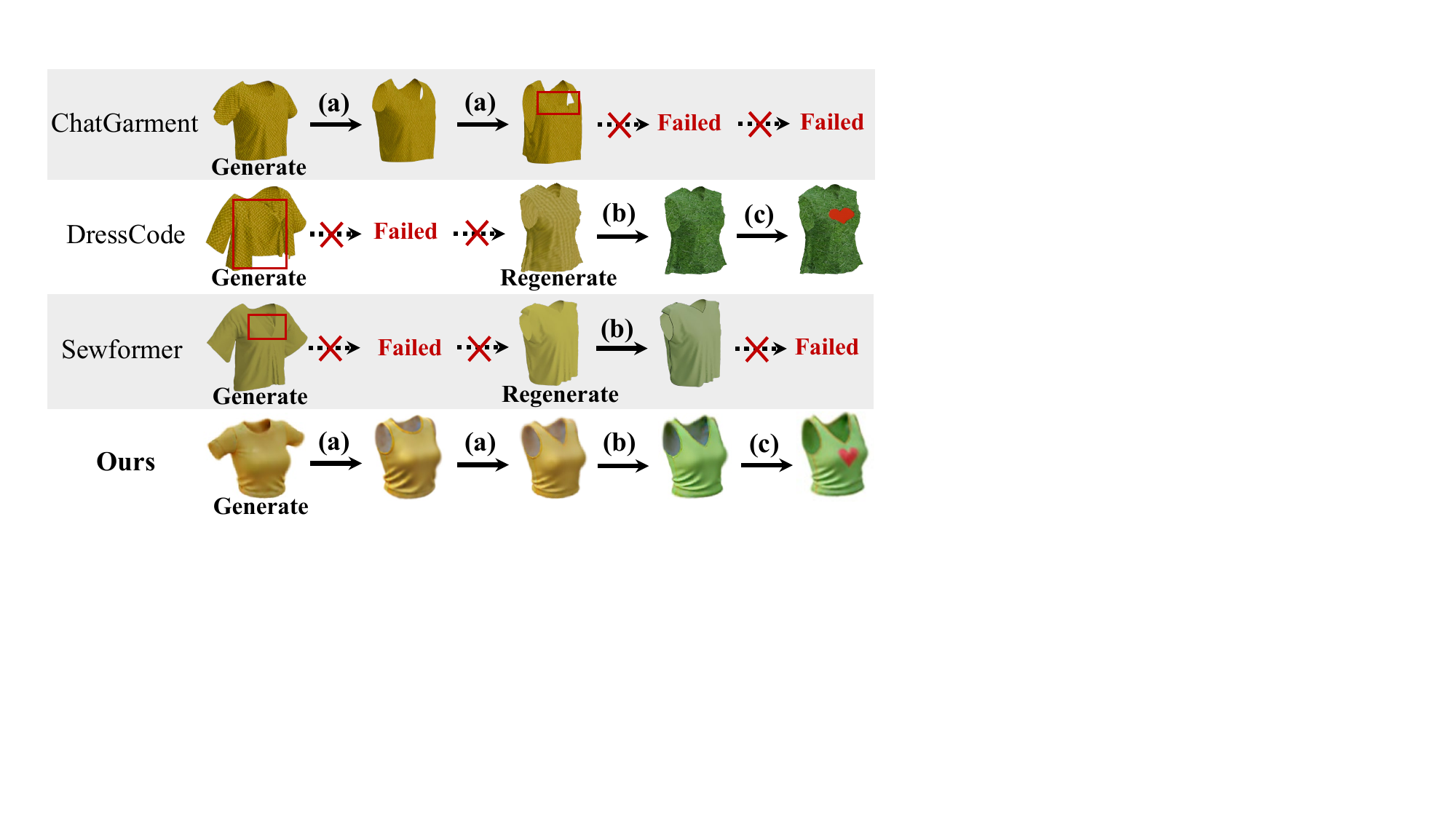}
    \caption{Qualitative comparisons on editing methods. In this figure, (a), (b), and (c) correspond to geometric editing, global texture editing, and local texture editing, respectively.}
\label{fig:edit_compare}
\end{figure}
Our framework is implemented based on Threestudio~\cite{guo2023threestudio} using PyTorch on a single RTX 4090 GPU, with batch size 4 and the size of the rendered images being 512×512. 
In the experiments, the initialized SMPL-X shape parameters \textbf{$\beta$} are set to zero, while the pose parameters \textbf{$\theta$} are typically set to zero (T-pose) or consistent with the pose of image prompts.
For 2D Diffusion guidance, the training iterations for clothing with text and image guidance are 800 and 2000, respectively, with guide scales of 10.0 and 3.0. 
In the self-occlusion optimization, both semantic optimization and smooth optimization are performed for a maximum of 1000 iterations.
The Gaussian learning rates for position \textbf{$\mu$}, scale \textbf{$s$}, color \textbf{$c$}, opacity $\alpha$, and rotation \textbf{$q$} are set to $5\times10^{-5}$, $5\times10^{-3}$, $10^{-2}$, $10^{-2}$, and $10^{-3}$, respectively. 
The densify gradient threshold is set to $2\times10^{-4}$, and the split threshold is 0.01.

\subsection{Results}
\subsubsection{Quantitative Comparison}
In Table.~\ref{table:Text-3d}, we use GPT-4~\cite{ouyang2022training} to randomly generate 10 appearance descriptions for each of 9 different categories of clothing and employ various methods to generate the garments. The rendered results from the front, back, left, and right views are selected for evaluation.
We employ three models, \textit{ViT-B/16}, \textit{ViT-B/32}, and \textit{ViT-L/14} from OpenAI\footnote{\href{https://huggingface.co/openai}{https://huggingface.co/openai}}, to compute the CLIP-Score~\cite{radford2021learning} for measuring the relevance between the clothing and the text prompt. 
We also conduct a user study to evaluate garments generated from textual descriptions, collecting a total of 240 responses from 20 participants.
Participants were asked to rate each method on a scale of 1 to 5, based on ``3D consistency'', ``text alignment with prompt'', and ``image quality'', which are three critical aspects in the 3D generation tasks. The result indicates that our method received the highest preference.
In Table.~\ref{table: Image-3d}, we rendered the DeepFashion3D v2 and ClothesNet datasets as ground truth, using the front view as the input for each method. We evaluated the methods based on image quality and similarity, assessing the PSNR, SSIM~\cite{wang2004image}, LPIPS~\cite{zhang2018unreasonable}, and CLIP Similarity~\cite{radford2021learning} metrics. 
The image-guided user study with 160 responses, evaluated the generated garments from image-guided methods based on multi-view consistency and image quality, with our method achieving the highest score.
By leveraging semantic initialization based on the SMPL-X prior and 2D diffusion guidance, our method demonstrates superior or comparable results in terms of text-image consistency, image quality, and 3D consistency. 
\subsubsection{Qualitative Comparison}
We provide qualitative comparisons with Shap-E~\cite{jun2023shap}, 3DGS~\cite{kerbl20233d} with StableDiffusion~\cite{rombach2022high}, LGM~\cite{tang2024lgm}, DreamGaussian~\cite{tang2023dreamgaussian}, GaussianDreamer~\cite{gaussiandreamer} and GarmentDreamer~\cite{li2024garmentdreamer} on text-guided garments and accessories in Fig.~\ref{fig:Text-3d}. 
Shap-E exhibits stable 3D consistency and color representation but lacks the detailed texture of the clothing. 3DGS, LGM, and GaussianDreamer effectively show fabric texture details, but suffer from blurry edges, color mismatches with the prompt, or the generation of unintended elements (e.g., human body). DreamGaussian produces clear clothing structures with vibrant colors, but the surfaces display unrealistic protrusions. We utilize the provided garment mesh template as input of GarmentDreamer, which generates highly realistic fabric simulations. However, the garment's design is constrained by the predefined mesh, limiting shape diversity and accessory generation. In contrast, our method generates high-quality clothing of diverse shapes that align with the text prompt.

For image input, we compare with Gaussian-based single-view reconstruction approaches across four views. 
As shown in Fig.~\ref{fig:Image-3d}, DreamGaussian exhibits strong multi-view consistency but suffers from a lack of smoothness on the garment surface. TriplaneGaussian~\cite{zou2024triplane}, Gamba~\cite{shen2024gamba}, and LGM are capable of generating realistic texture details; however, they exhibit noticeable garment flattening or garment tilting to some extent in side views. 
In contrast, our method consistently produces anatomically accurate results across all views, obviating the need for manual adjustments and enabling more efficient applications. 
\begin{figure}[!tb]
\setlength{\abovecaptionskip}{4pt} 
\centering
    \includegraphics[width=0.9\linewidth]{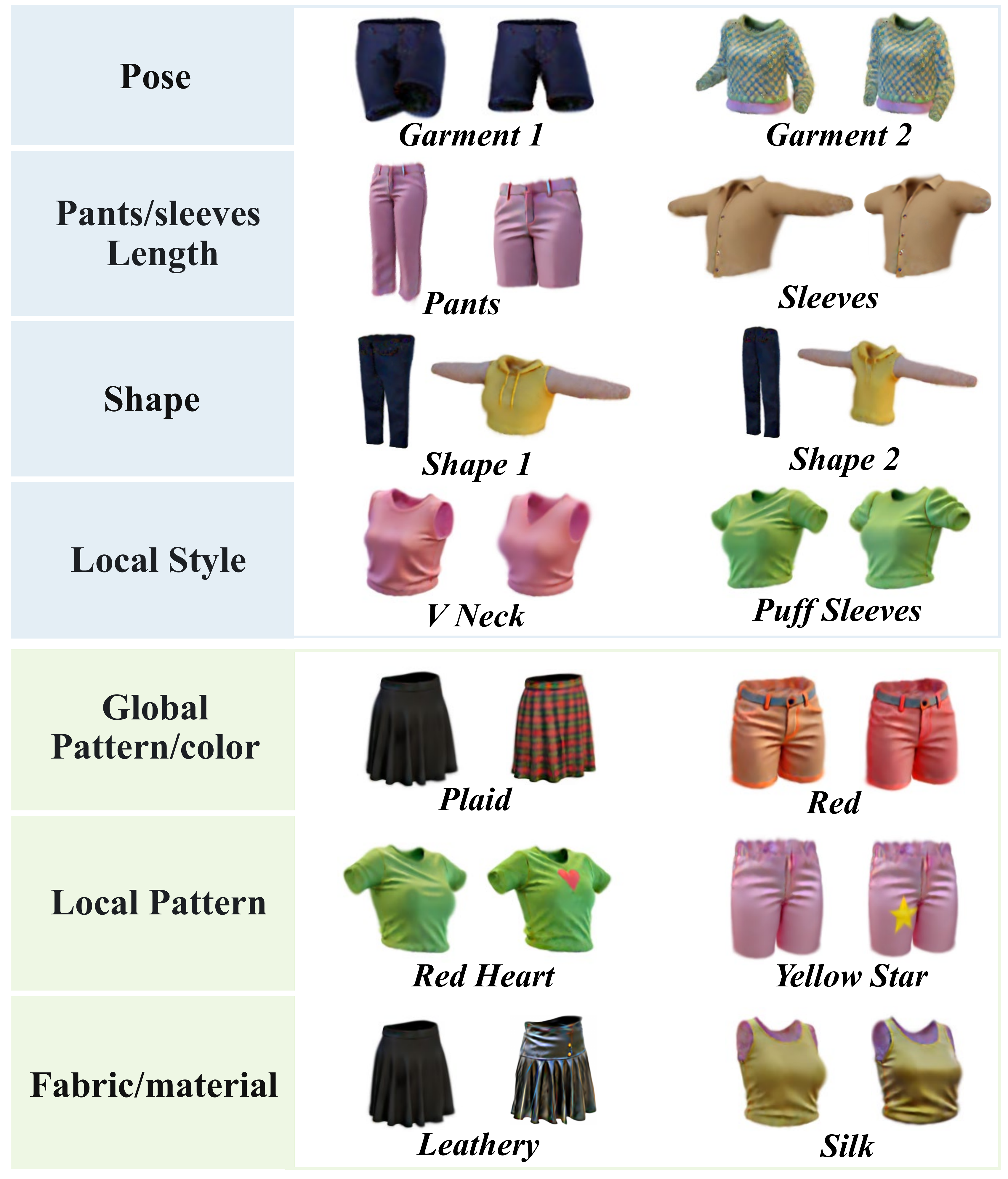}
    \caption{Results of editing garments. Green and blue color correspond to texture-related and geometry-related editing.}
\label{fig:edit}
\end{figure}
Figure.~\ref{fig:edit_compare} presents the qualitative comparison with three garment editing methods: ChatGarment~\cite{bian2024chatgarment}, DressCode~\cite{he2024dresscode}, and Sewformer~\cite{liu2023towards}. We aim for each method first to generate a yellow round-neck T-shirt, followed by geometric edits to produce a yellow round-neck vest and then a V-neck vest. Subsequently, a global texture edit is applied to obtain a green V-neck vest. Finally, a local texture edit is performed on the chest pattern. ChatGarment supports stable geometric editing but lacks texture editing capabilities, while DressCode and Sewformer exhibit limitations in geometric editing. The results of other comparison methods are all rendered using Blender~\cite{community2018blender}, and ChatGarment uses the texture generated by DressCode. The results demonstrate that our method enables high-quality and progressive edits.

\subsubsection{Garment Editing}
Figure.~\ref{fig:edit} demonstrates that our method can meet various editing requirements, such as material, shape, pose and style editing. Owing to the clothing semantic model, our approach not only supports global modifications but also enables texture and geometric editing in desired regions.
The generated garments can directly fit various body shapes and poses without the need for software such as Blender~\cite{community2018blender}.
\subsubsection{Accessories Wearing}
Our method can generate different accessories based on human semantic priors, such as shoes, glasses, and hats. More video results can be found at our \href{https://semanticgarment.github.io/SemanticGarment/}{project page}.

\subsection{Ablation Study}
\begin{figure}[!tb]
\setlength{\abovecaptionskip}{4pt} 
\centering
\vspace{-7pt}
\includegraphics[width=1\linewidth]{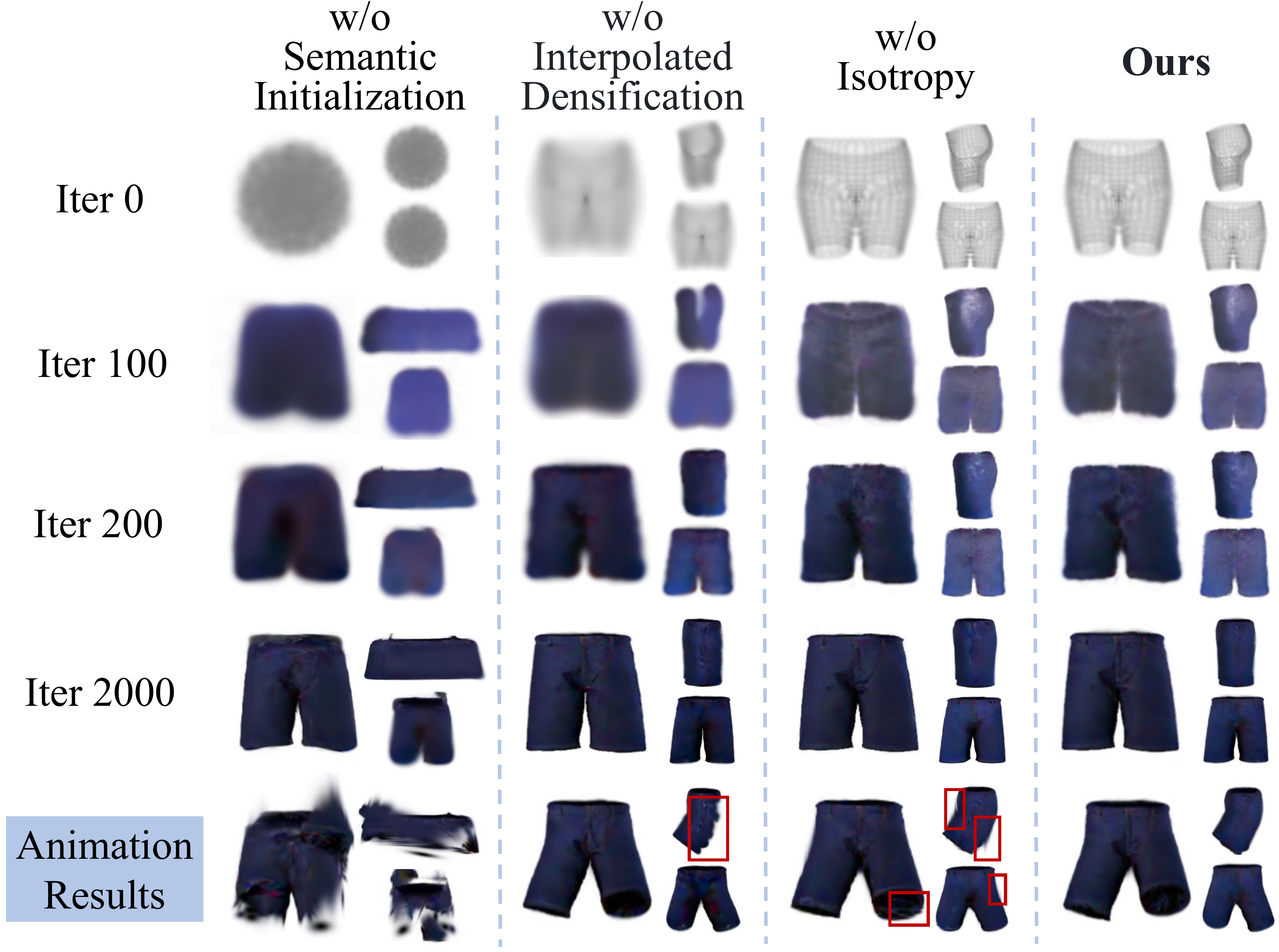}
    \caption{Ablation studies on semantic initialization, interpolated densification strategy and the isotropy of Gaussians.}
\label{fig:abla_gs}
\end{figure}
We conduct ablation studies on 3D Gaussian in garment generation, and the effectiveness of semantic editing, self-occlusion optimization and semantic optimization.
The results shown in Fig.~\ref{fig:abla_gs}, Fig.~\ref{fig:abla_edit}, Fig.~\ref{fig:abla_so} and Fig.~\ref{fig:abla_sematic} demonstrate that the absence of any of these designs results in a decline in quality.
 
\subsubsection{3D Gaussians}
\noindent\textbf{1) Semantic initialization.}
The results in the first column of Fig.~\ref{fig:abla_gs} are obtained by randomly initializing the Gaussian points without semantic initialization in Sec.~\ref{sec:semantic_initialization_prior}. From the side view, it is evident that the shorts exhibit incorrect thickness and produce entirely distorted driving results.
\noindent\textbf{2) Interpolated densification.}
In the second column, the results are obtained without using our interpolated densification strategy, leading to geometric instability in the driving results.
\noindent\textbf{3) Isotropy of Gaussians.} 
The outcomes in the third column are from experiments in which the Gaussian points were not isotropic, resulting in blurry Gaussian artifacts in animation across different views.

\subsubsection{Semantic Edit}
\begin{figure}[!tb]
\setlength{\abovecaptionskip}{4pt} 
\centering
\includegraphics[width=0.9\linewidth]{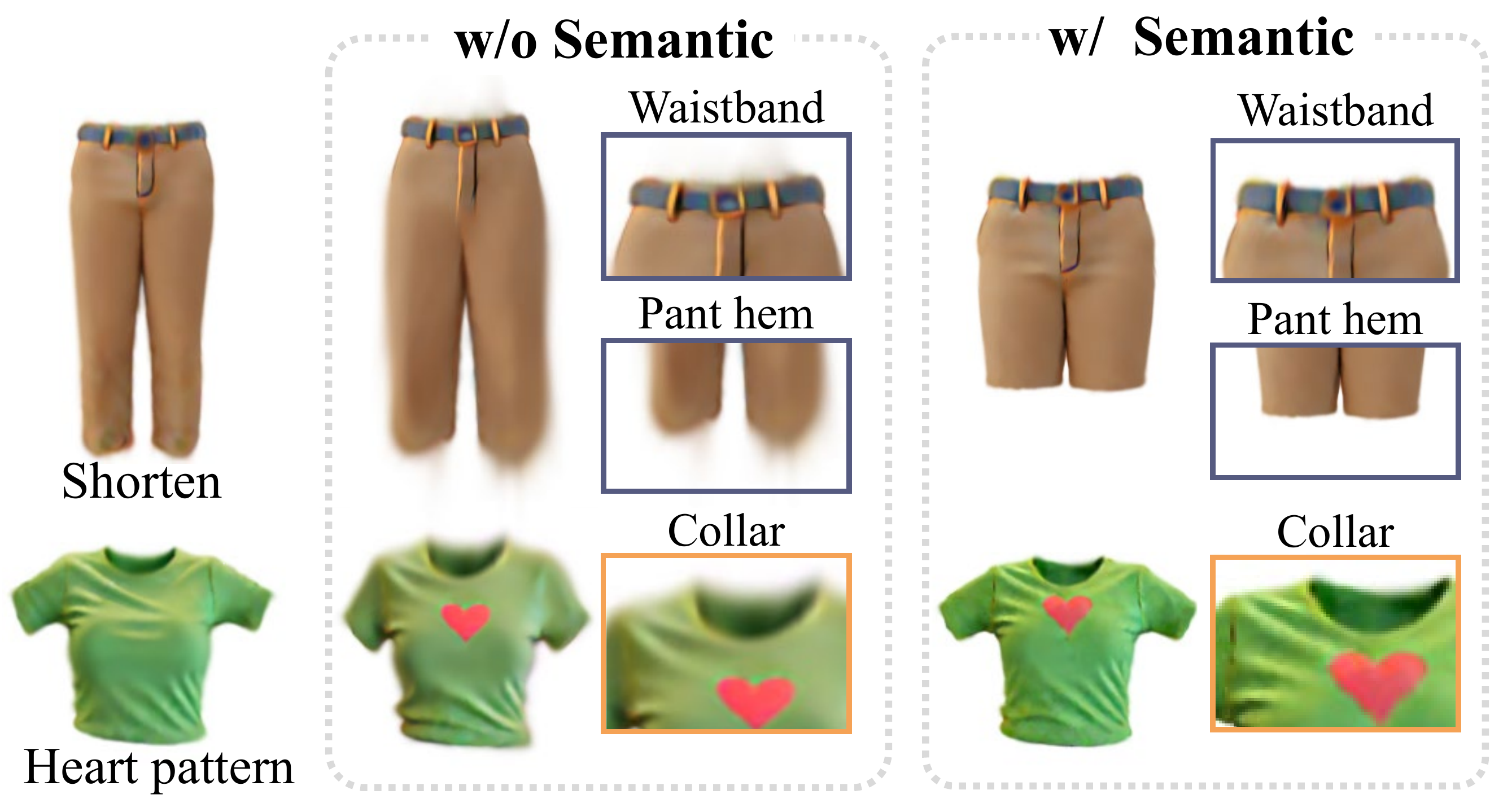}
    \caption{Ablation studies on semantic editing. With the assistance of semantic information, we can restrict the editing region to the desired area and avoid blurry boundaries.}
\label{fig:abla_edit}
\end{figure}
As shown in Fig.~\ref{fig:abla_edit}, solely relying on the SDS loss tends to modify the entire garment and blurring at the edges. In contrast, utilizing semantic information enables flexible pruning and densification of the target region, better preserving the color and shape of the remaining areas.

\subsubsection{Self-Occlusion Optimization}
\begin{figure}[!tb]
\setlength{\abovecaptionskip}{4pt} 
\centering
\includegraphics[width=0.9\linewidth]{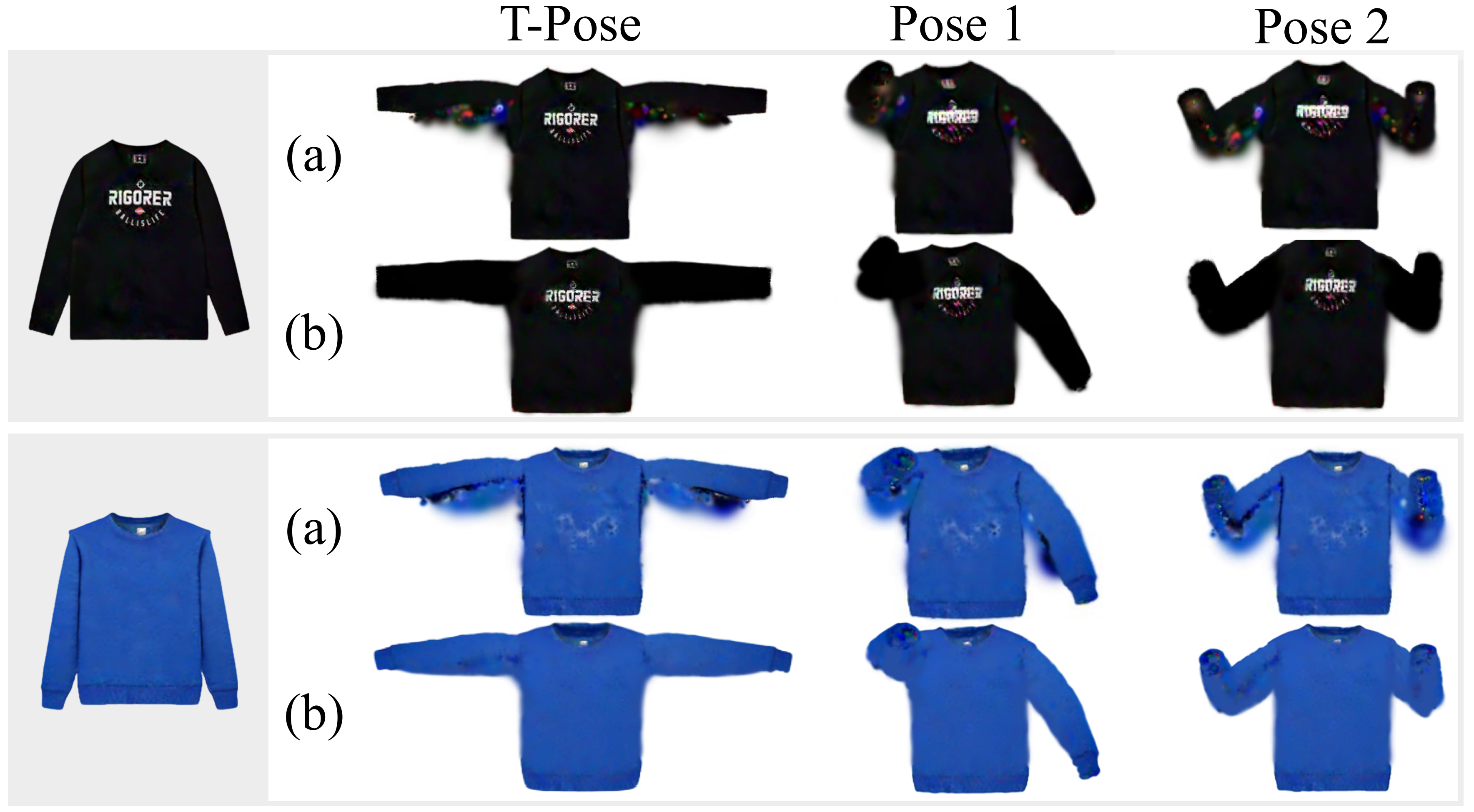}
    \caption{Ablation studies on self-occlusion optimization. For each garment, (a) and (b) show the results without and with self-occlusion optimization, respectively.}
\label{fig:abla_so}
\end{figure}
In Fig.~\ref{fig:abla_so}, we present the ablation results of self-occlusion optimization in different driving poses. The results demonstrate that incorporating self-occlusion optimization significantly improves the shape of self-occluded regions and reduces artifacts.

\subsubsection{Semantic Optimization}
\begin{figure}[!tb]
\centering
\setlength{\abovecaptionskip}{4pt} 
\includegraphics[width=1\linewidth]{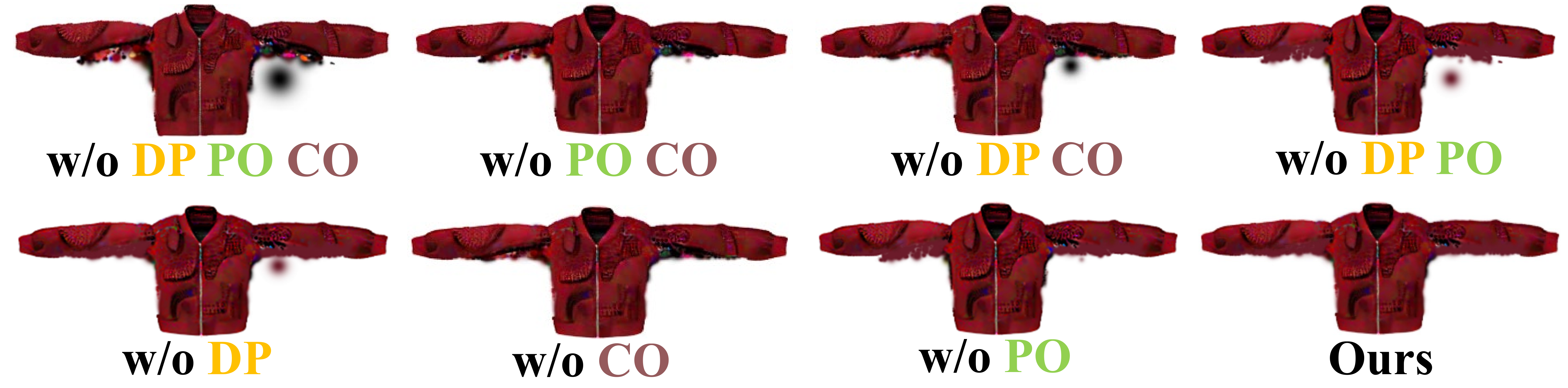}
    \caption{Ablation studies on semantic optimization. ``DP'', ``PO'' and ``CO'' refer to distance-based pruning, position optimization, and color optimization, respectively.}
\label{fig:abla_sematic}
\end{figure}

In Fig.~\ref{fig:abla_sematic}, we assess the impact of three components in semantic optimization in Sec.~\ref{sec:semantic_optimization}. The distance-based pruning efficiently eliminates points that are far from the garment. Position optimization facilitates the movement of Gaussians closer to the garment surface. Color optimization helps mitigate the impact of pose changes on the color of self-occluded regions.

\section{Conclusion}
\label{sec:conclusion}
In this paper, we propose SemanticGarment, a 3D garment Gaussian generation and editing method guided by bi-modal prompts (image or text) with strong semantic priors.
We first introduce a high-fidelity 3D semantic clothing model encompassing semantic information of diverse garments, accessories and key regions to achieve more flexible and accurate control over 3D digital garments.
To avoid the resource consumption of regeneration, we present a semantic-aware editing approach, enabling global or targeted area modifications, including texture, material, shape, and pose.
Furthermore, SemanticGarment designs a self-occlusion optimization strategy that enables the completion and refinement of color and shape in occluded garment areas. 
These capabilities not only enhance automation in garment design but also open new possibilities for personalized clothing customization and immersive virtual try-on experiences.

\noindent\textbf{Limitations and Future work.}
Although SemanticGarment demonstrates promising results, it still has several limitations. 
First, the texture or shape guided by text relies on pretrained diffusion guidance, which may perform poorly when encountering complex prompts.
Second, using the \textit{LBS} function to drive loose clothing (e.g., skirts) is less effective compared to fitted clothing. In future work, we plan to explore extracting the mesh structure from the generated Gaussians and applying physics-based simulations for animation. 
\begin{acks}
This work was partly supported by the NSFC62431015, Science and Technology Commission of Shanghai Municipality No.24511106200, the Fundamental Research Funds for the Central Universities, Shanghai Key Laboratory of Digital Media Processing and Transmission under Grant 22DZ2229005, 111 project BP0719010.
\end{acks}

\bibliographystyle{ACM-Reference-Format}
\balance
\bibliography{sample-base}

\end{document}